\documentstyle[10pt,emulateapj]{article}

\received{23 August 1999}
\accepted{30 November 1999}

\begin{document}

\title{Baryon Distribution in Galaxy Clusters as a Result of  
       Sedimentation of Helium Nuclei}

\author{Bo Qin and Xiang-Ping Wu}
\affil{Beijing Astronomical Observatory, Chinese Academy 
                 of Sciences, Beijing 100012, China} 
\affil{National Astronomical Observatories, Chinese Academy 
                 of Sciences, Beijing 100012, China}
\submitted{ApJ Letters, 2000, 529, L1}

\begin{abstract}

Heavy particles in galaxy clusters tend to be more centrally concentrated 
than light ones according to  the Boltzmann distribution.
An estimate of the drift velocity suggests that it is possible  
that the helium nuclei may have entirely or partially sedimented 
into the cluster core within the Hubble time.
We demonstrate the scenario using the Navarro-Frenk-White profile as 
the dark matter
distribution of clusters and assuming that the intracluster gas is 
isothermal and in hydrostatic equilibrium. 
We find that a greater fraction of baryonic matter is distributed at 
small radii than at large radii, which challenges the prevailing claim 
that the baryon fraction increases monotonically with cluster radius.
It shows that the conventional mass estimate using X-ray measurements
of intracluster gas along with a constant mean molecular weight 
may have underestimated the total cluster mass
by $\sim 20\%$, which in turn leads to an overestimate of the 
total baryon fraction by the same percentage.
Additionally, it is pointed out that the sedimentation of helium nuclei
toward cluster cores may at least partially account for the sharp peaks 
in the central X-ray emissions observed in some clusters.
\end{abstract}


\section{Introduction}

An accurate determination of the total gravitating mass of galaxy clusters 
is crucial for the `direct' measurements of the cosmic mass density
parameter $\Omega_M$ through the mass-to-light technique 
(e.g. Bahcall, Lubin \& Dorman 1995) and the baryon
fraction method (e.g. White et al. 1993). The latter has received 
much attention in recent years because of the rapid progress in
X-ray astronomy and particularly the large spatial extension 
of the hot and diffuse X-ray emitting gas in clusters. 
In the conventional treatment, the volume-averaged baryon fraction $f_b$,
defined as the ratio of the gas mass $M_{\rm gas}$ to the total mass 
$M_{\rm tot}$
of a cluster, is obtained by assuming a thermal bremsstrahlung emission
and hydrostatic equilibrium for the intracluster gas.
While such an exercise has been done for almost every X-ray selected
cluster with known temperature, the resultant baryon fractions show
rather a large dispersion among different clusters. In particular,
$f_b$ appears to be a monotonically increasing function of cluster radius, 
and thereby cannot, in principle, be used for the determination of 
the cosmic density parameter  $\Omega_M$ since the asymptotic form of
$f_b$ at large radii does not approach a universal value 
(White \& Fabian 1995; David et al. 1995; Markevitch \& Vikhlinin 1997a,b; 
White et al. 1997; Ettori \& Fabian 1999; Nevalainen et al. 1999;
Markevitch et al. 1999; Wu \& Xue 1999). This leads to an uncomfortable
situation that a greater fraction of dark matter is distributed 
at small scales than at large scales. 
It is commonly believed that these puzzles have arisen from our 
poor understanding of the local dynamics inside clusters
such as cooling/non-cooling flows, substructures, mergers, etc. 
Yet, a satisfactory explanation has not been achieved.

The conventional cluster mass estimate from X-ray measurement
of intracluster gas assumes a constant mean molecular weight $\mu$.
That is, all the chemical elements of the X-ray emitting gas  
obey exactly the same spatial distribution.     
Under this assumption, the total mass of the cluster
$M_{\rm tot} \propto \mu^{-1}$, while the mass in gas 
$M_{\rm gas} \propto \mu$. Consequently, the cluster baryon fraction 
$f_b (\approx \! M_{\rm gas} / M_{\rm tot})$
depends sensitively on the value of $\mu$. 
It appears that our estimates of total mass and baryon fraction of a 
cluster would be seriously affected 
if its chemical elements do not share the same spatial distribution.

The Boltzmann distribution of particles in a gravitational field $\phi$
follows $n \propto \exp\,(-m\phi/kT)$, where $n$ and $m$ are the number 
density and the individual mass of the particles, respectively. 
The heavier a particle is, the slower its thermal motion 
will be. So, heavy ions in the intracluster gas
will have a tendency to drift toward the cluster center.
As a consequence of this sedimentation of heavy ions toward the
central region, given that the cluster has survived  
for a sufficiently long time in the Universe,  
$\mu$ will no longer be a constant in the cluster.
If the intracluster gas is entirely composed of hydrogen and
helium with their primordial abundances,  
the value of $\mu$ will be higher at cluster center 
while asymptotically reaches 0.5 at large radii
for a fully ionized hydrogen gas.

The chemically inhomogeneous distribution in galaxy clusters was  
initiated by Fabian \& Pringle (1977), who studied the sedimentation 
of iron nuclei. They found that the iron nuclei 
in the X-ray gas may settle into the cluster core within a Hubble time. 
Taking into account the collisions of iron nuclei with helium nuclei,
Rephaeli (1978) argued that a much longer sedimentation time than the Hubble 
time was required so that the iron nuclei would not have 
enough time to sediment into the core. 
In this {\it Letter} we wish to focus on helium nuclei instead of iron nuclei.
Since the drift velocity $v_D \propto AZ^{-2}$ (e.g. Fabian \& Pringle 1977), 
where $A$ and $Z$ are the atomic weight and the charge of 
the ion, respectively, the helium nuclei will settle much faster than 
the iron nuclei, and thus the sedimentation may eventually take place.
Indeed,  Abramopoulos, Chana \& Ku (1981) have calculated the equilibrium 
distribution of the elements in the Coma cluster, assuming an 
analytic King potential, and found that helium and other heavy 
elements are strongly concentrated to the cluster core. In particular, 
Gilfanov \& Sunyaev (1984) have demonstrated that the diffusion of 
elements in the X-ray gas may significantly enhance the deuterium, helium,
and lithium abundances in the core regions of rich clusters of galaxies.
For simplicity, we assume that the intracluster gas consists of 
hydrogen and helium, and then demonstrate  their equilibrium distributions 
in the gravitational potential characterized by the universal density profile 
(Navarro, Frenk \& White 1995, 1997; hereafter NFW). 
As will be shown, under the scenario of helium sedimentation,  
the baryonic matter in clusters 
can be more centrally distributed than the dark matter, in 
contradiction to what is commonly believed. This may open a possibility 
to solve the puzzle for an increasing baryon fraction with cluster radius.

\section{Sedimentation of helium nuclei in galaxy clusters}

Our working model is based on the assumption that the intracluster gas 
composed of hydrogen and helium is in 
hydrostatic equilibrium at a temperature $T$ with the underlying
gravitational potential.  This can be justified by a simple estimate of
their relaxation time $t_{\rm eq}$ (Spitzer 1978)
\begin{eqnarray}
t_{\rm eq} & = &
	\frac {3 m_p m_{\alpha} k^{3/2}} 
	      {8(2\pi)^{1/2} n_p Z_p^2 Z_{\alpha}^2 e^4 \ln {\Lambda}}
        \left( \frac{T}{m_p} + \frac{T}{m_{\alpha}}  \right)^{3/2} 
	\nonumber \\
       & = &
	6 \times 10^6 {\rm yr} \left( \frac{n_p} {10^{-3}{\rm cm}^{-3} }
					\right)^{-1}
			       \left( \frac{T} {10^{8}{\rm K}}
					\right)^{3/2},	
\end{eqnarray}
where $m_p$, $m_{\alpha}$, $Z_p$, and $Z_{\alpha}$ are the masses and the
charges of proton and helium nucleus, respectively. $n_p$ is the proton 
number density, and the Coulomb logarithm is taken to be  
$\ln {\Lambda} \approx 40$. 
Obviously, $t_{\rm eq}$ is much shorter than the present age of the Universe, 
and hence, the protons and the helium nuclei are readily 
in hydrostatic equilibrium at the same temperature.

According to the Boltzmann distribution of particles in a gravitational 
field, heavy particles tend to be more centrally distributed in 
galaxy clusters than light particles. Therefore, 
with respect to protons, helium nuclei will tend to drift toward 
the cluster center. An immediate question is: Is the drift velocity 
sufficiently large for the helium nuclei to have settled into the 
cluster core within the Hubble time ? 
The drift velocity $v_D$ of helium nuclei  can be estimated through 
(Fabian \& Pringle 1977)
\begin{eqnarray}
v_D & = 
	& \frac {3 A_{\alpha} m_p^{1/2} g (2kT)^{3/2} }
		{16 \pi^{1/2} Z_{\alpha}^2 e^4 n_p \ln {\Lambda} } 
			\nonumber \\
 & = &  8.8 \times 10^6 {\rm cm/s} 
                    \left( \frac{g}{3 \times 10^{-8} {\rm cm\, s}^{-2} } 
                                             \right)  \nonumber \\
                  &   & \times               
                    \left( \frac{n_p}{10^{-3}{\rm cm}^{-3} } \right)^{-1}  
                    \left( \frac{T}{10^8 {\rm K}} \right)^{3/2}, 
\end{eqnarray}
where $g$ is the gravitational acceleration. Eq.(2) indicates that within 
the Hubble time the helium nuclei can drift a distance
\begin{equation}
r_D \! = \! 1.8 \, {\rm Mpc} \!
	     \left( \! \frac{g}{3 \! \times \! 10^{-8} {\rm cm\, s}^{-2} } 
			\! \right) \!\!
             \left( \! \frac{n_p}{10^{-3}{\rm cm}^{-3} } 
			\! \right)^{\!\!-1} \!\!\!  
             \left( \! \frac{T}{10^8 {\rm K}} \! \right)^{\!\!3/2} \!\!\!
		    h_{50}^{-1},	 
\end{equation}
which is indeed comparable to cluster scales. Moreover, as 
$r_D \propto g \, n_p^{-1}$, the value of
$r_D$ (and $v_D$) increases rapidly with cluster radius $r$. 
Therefore, Eq.(3) suggests that the majority of the helium nuclei 
have probably sedimented into the cluster core within the Hubble time. 
Note that, due to the requirement of electrical neutrality, 
the electrons of the same charge will simultaneously sediment 
along with the helium nuclei. Here, we have not considered
the effects of magnetic fields and subcluster mergers, which
may somewhat retard the sedimentation of helium nuclei  
(Rephaeli 1978; Gilfanov \& Sunyaev 1984).

The sedimentation of helium nuclei in galaxy clusters will lead to 
a dramatic change of baryonic matter distribution. Consequently,
the determination of gas and total mass of a cluster will be affected
through the mean molecular weight $\mu$. Another significant 
effect is that a sharp peak in the X-ray emission concentrated in 
cluster core will be expected due to the electron - helium nucleus radiation.
This provides an alternative scenario for the `cooling flows' seen in
some clusters, for which a detailed investigation will be presented 
elsewhere.  In the present {\it Letter},  we only focus on the 
dynamical effect. In the extreme case, the X-ray gas will be 
helium-dominated at cluster center, while hydrogen-dominated at 
large radii. The mean molecular weight $\mu$, which is commonly 
used as a constant, will be a decreasing function of the cluster
radius. At cluster center, $\mu$ reaches $4/3$, the value for a fully 
ionized helium gas, while at large radii, $\mu$ approaches $0.5$, 
the value for a fully ionized hydrogen gas.

The conventional mass estimate from X-ray measurement assumes a 
constant mean molecular weight of 
$\mu = 0.59$. As the {\it total} dynamical mass $M_{\rm tot}$ 
of a cluster is uniquely determined by the intracluster gas 
at large radii, the difference between $M_{\rm tot}^c$ 
(the value from the conventional method with $\mu = 0.59$) 
and $M_{\rm tot}$ as a result of helium sedimentation is simply  
\begin{equation}
M_{\rm tot} = \frac{0.59}{0.5} M_{\rm tot}^c
	    = 1.18 M_{\rm tot}^c.
\end{equation}
This indicates that the conventional method using X-ray measurement
of intracluster gas, together with a constant mean molecular weight,
may have underestimated the total cluster mass by $\sim 20\%$, which 
in turn, results in an overestimate of the total baryon fraction by 
the same percentage. While the effect of the helium concentration toward 
cluster center can alter the gas distribution,
the total mass in gas of the {\it whole}
cluster remains unaffected because of the mass conservation.

\section{Gas distribution under the NFW potential}

We now demonstrate how hydrogen and helium are distributed in
clusters described by the NFW profile
\begin{equation}
\rho = \frac {\rho_s}
	     { (r/r_s) (1+r/r_s)^2 }.
\end{equation}
As has been shown by Makino et al. (1998), such a potential results in
an analytic form of gas number density:
\begin{equation}
n_{\rm gas}(x) = n_{\rm gas}(0) \, e^{-\eta_{\rm gas}} 
			(1+x)^{\eta_{\rm gas}/x}, 
\end{equation}
where $x=r/r_s$, and 
$\eta_{\rm gas} = 4\pi G \mu m_p \rho_s r_s^2 /kT$.
Except for the small core radius, $n_{\rm gas}(x)$ is well approximated 
by the conventional $\beta$ model. If we neglect the interaction between 
protons and helium nuclei, then 
\begin{equation}
n_p = n_{p0} \, e^{-\eta} (1+x)^{\eta/x}
\end{equation}
\begin{equation}
n_{\alpha} = n_{\alpha 0} \, e^{-8\eta/3} (1+x)^{8\eta/3x}
\end{equation}
where $\eta = 2\pi G m_p \rho_s r_s^2 /kT$. Eqs (7) and (8) give
$n_{\alpha} \propto n_p^{8/3}$, which differs from the prediction 
by Gilfanov \& Sunyaev (1984), $n_{\alpha} \propto n_p^6$.
The discrepancy is probably due to the fact that we have not accounted
for the diffusion-induced electric fields.

We display in Fig.1 the radial distributions of protons and helium nuclei
as well as their combined result for a typical nearby cluster with
$kT=7$ keV, $\eta_{\rm gas}=10$ and $r_s=1$ Mpc (e.g. Ettori \& Fabian 1999).
It is apparent that the intracluster gas is dominated by different elements
at different radius ranges: Within the core radius of a few tenth of $r_s$,
the number density of helium is about four times larger than 
that of hydrogen because of the sedimentation of helium nuclei,  
giving rise to a significant excess of both mass in gas
(see Fig.2) and X-ray emission in the central region of the cluster
relative to the conventional model. Outside the core radius, 
helium profile shows a sharp drop and protons become 
the major component of the gas. It has been claimed that the total 
gas distribution can be approximated by the $\beta$ model 
(Makino et al. 1998). In fact, it is easy to show that both the narrow (helium)
and extended (hydrogen) components of intracluster gas can be fitted by
the $\beta$ models with different $\beta$ parameters. This provides a natural
explanation for the double $\beta$ model advocated recently for 
the cooling flow clusters
(e.g. Ikebe et al. 1996; Xu et al. 1998; Mohr, Mathiesen \& Evrard 1999; etc.).
Also plotted in Fig.1 is the mean molecular weight calculated by 
$\mu=(n_p+4n_{\alpha})/(2n_p+3n_{\alpha})$. The constant mean molecular
weight is now replaced by a decreasing function of radius. 
The asymptotic values of $\mu$ at small and large radii are 
$4/3$ and $1/2$, respectively. 

\placefigure{fig1}

We present in Fig.2 a comparison of gas masses determined 
by the invariant and variant mean molecular weights using the same cluster
parameters as in Fig.1. To facilitate the comparison, we require that 
the total particles within $r_{200}$ remains unchanged,
where $r_{200}$ is the radius within which the mean cluster mass density
is 200 times the critical mass density of the Universe ($\Omega_0=1$).
It appears that the sedimentation of helium nuclei 
leads to a remarkable concentration of baryonic matter towards cluster 
center, which challenges the conventional prediction that the baryon
fraction increases monotonically with radius (Fig.3).  
This opens a possibility that the asymptotic baryon fraction at large radii
may match the universal value defined by the Big Bang Nucleosysthesis,
although a detailed investigation will still be needed.

\placefigure{fig2}
\placefigure{fig3}

\section{Conclusions}

Intracluster gas is mainly composed of hydrogen and helium. Their average 
abundances over a whole cluster should be of the cosmic mixture. However,
their spatial distributions are entirely determined by the underlying
gravitational potential of the cluster, and thus follow the 
Boltzmann distribution. On the other hand, 
clusters are believed to have formed at redshift 
$z\ga 1$. Therefore, the helium nuclei in clusters may have entirely or
partially sedimented into the central cores of clusters today. This will 
lead to a significant change of the radial distributions of gas and 
baryon fraction. In the present {\it Letter}, we have only discussed 
the impact on the dynamical aspect of clusters. We will present elsewhere the 
effect on the cluster X-ray emission (e.g., cooling flows, the double
$\beta$ model, etc.) as a result of the sedimentation of helium nuclei.

In the conventional treatment where the mean molecular weight is assumed to
be constant, one may have underestimated the total dynamical mass of 
clusters by $\sim 20\%$. Using a more vigorous way in which the 
NFW profile is taken as the
background gravitational field of clusters,  we have studied the hydrogen
and helium distributions. Indeed, the sedimentation of helium nuclei toward
cluster centers has significantly changed the distribution of intracluster
gas, with gas being more centrally concentrated than dark matter. This may 
open a possibility to resolve the puzzle that the baryon fraction increases 
monotonically with radius predicted by the conventional model. 
In a word,  a number of cosmological applications of the dynamical properties 
of clusters will be affected by the sedimentation of helium nuclei if
it has really taken place during the evolution of clusters, which includes
the determination of $\Omega_M$ through $f_b$, the constraints on
the cosmological models through X-ray luminosity - temperature relation,
etc.  A detailed theoretical study, together with the observational 
constraints,  will be made in subsequent work.

\acknowledgments

We thank an anonymous referee for helpful comments.
This work was supported by the National Science Foundation of China, 
under Grant No. 1972531.




\figcaption{Distribution of intracluster gas tracing the gravitational
potential by the NFW profile for a typical nearby cluster, $kT=7$ keV, 
$\eta_{gas}=10$ and $r_s=1$ Mpc. The saturated distributions of hydrogen
and helium as well as their combined result 
are shown by dotted lines and solid line, respectively. All the 
profiles are scaled by the central number density of hydrogen. 
The corresponding mean molecular weight is displayed 
in the bottom panel by solid line. Also plotted is 
the conventionally adopted value $\mu=0.59$ (dotted line). 
\label{fig1}}

\figcaption{Different mass components plotted against cluster radius
for the same cluster in Fig.1. The dashed line shows the mass in gas
obtained in the conventional model, in which the central gas 
density is assumed to be $10^{-3}$ cm$^{-3}$. 
The conservation of total 
baryonic particles (or mass) within a radius of $r_{200}=3.1$ Mpc 
is required  for both with and without the sedimentation of helium nuclei. 
\label{fig2}}

\figcaption{A comparison of baryon fractions between the conventional model
(dotted line) and the scenario with the sedimentation of helium nuclei
(solid line). The cluster parameters are the same as in Fig.2.
\label{fig3}}

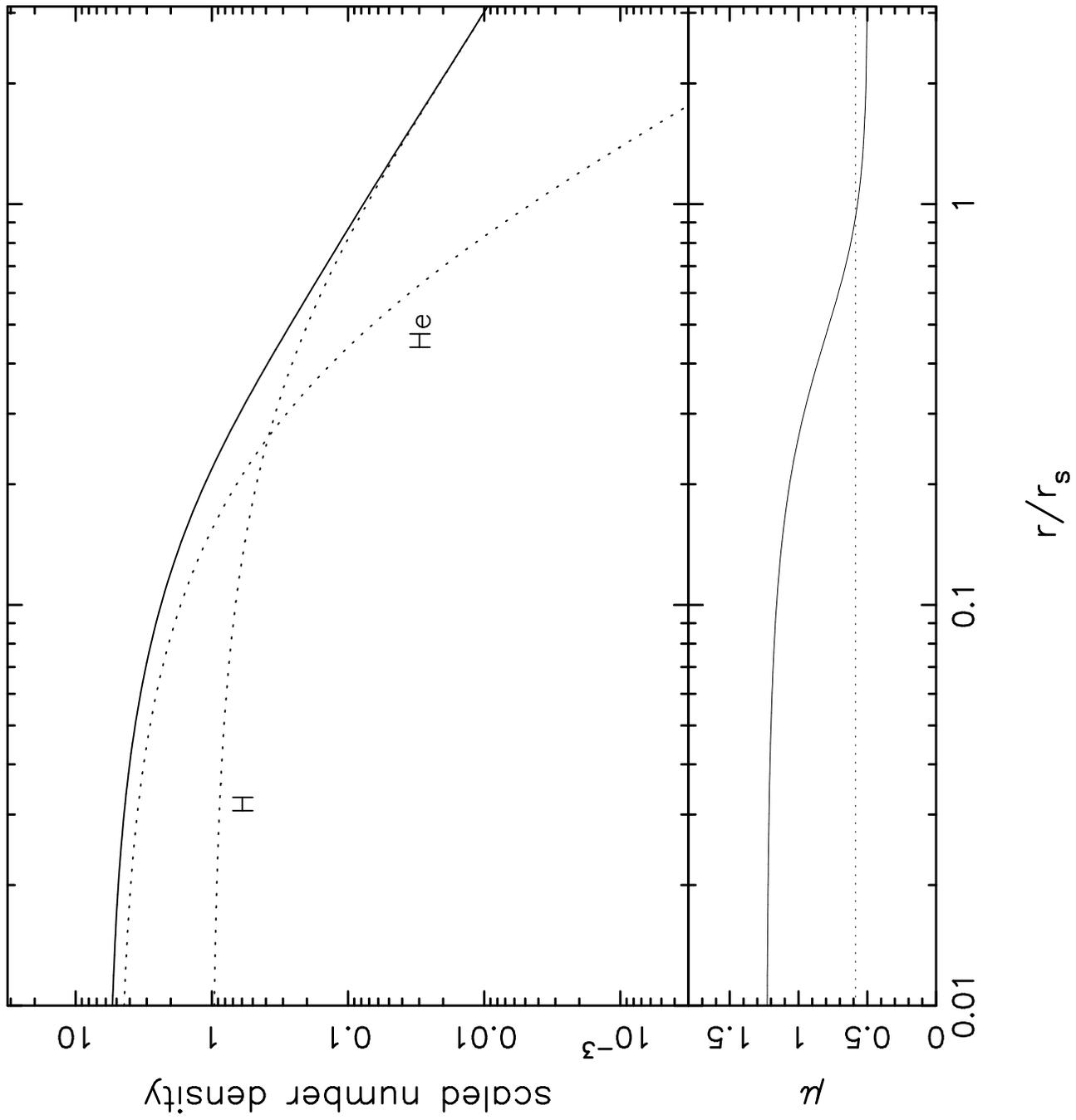
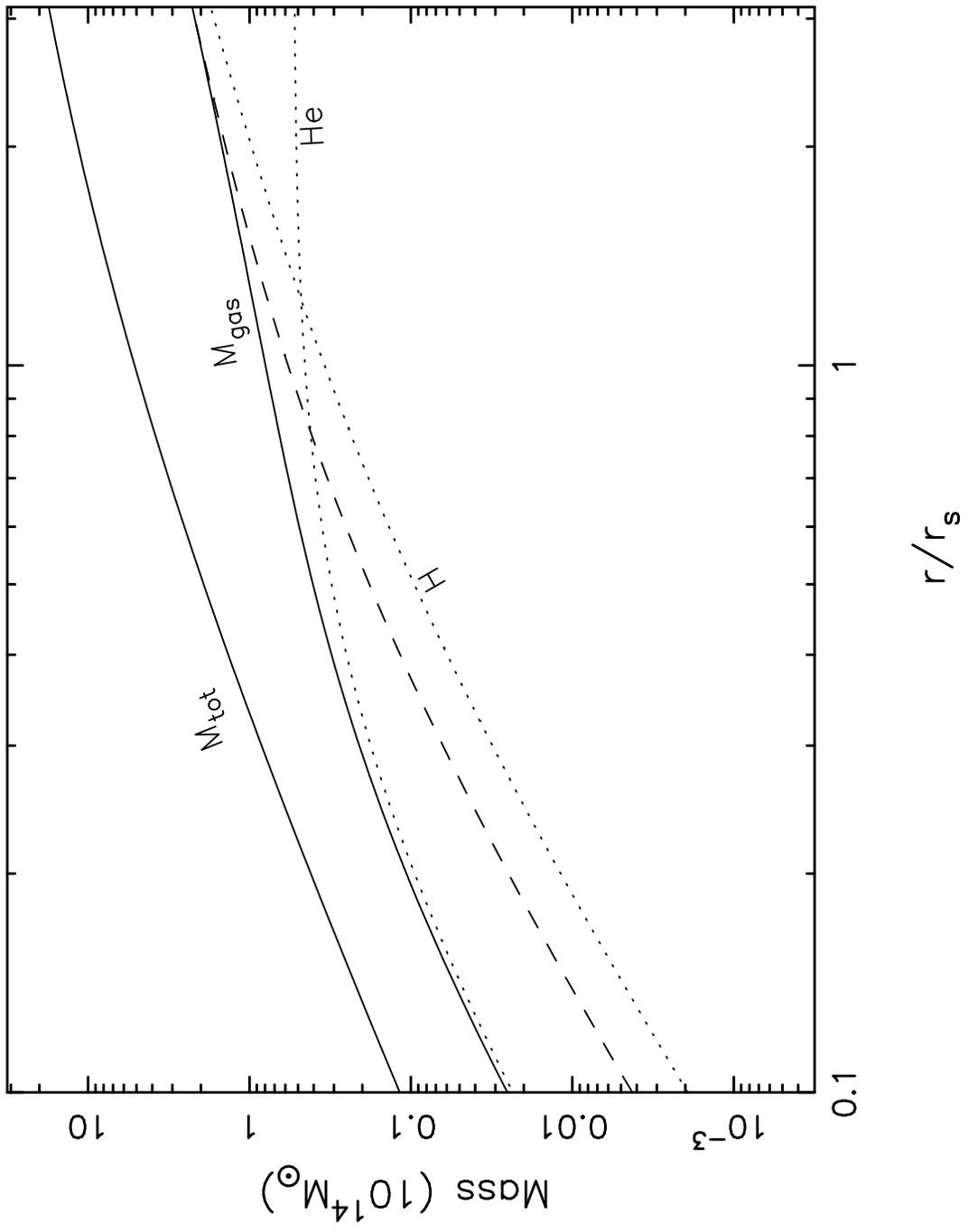
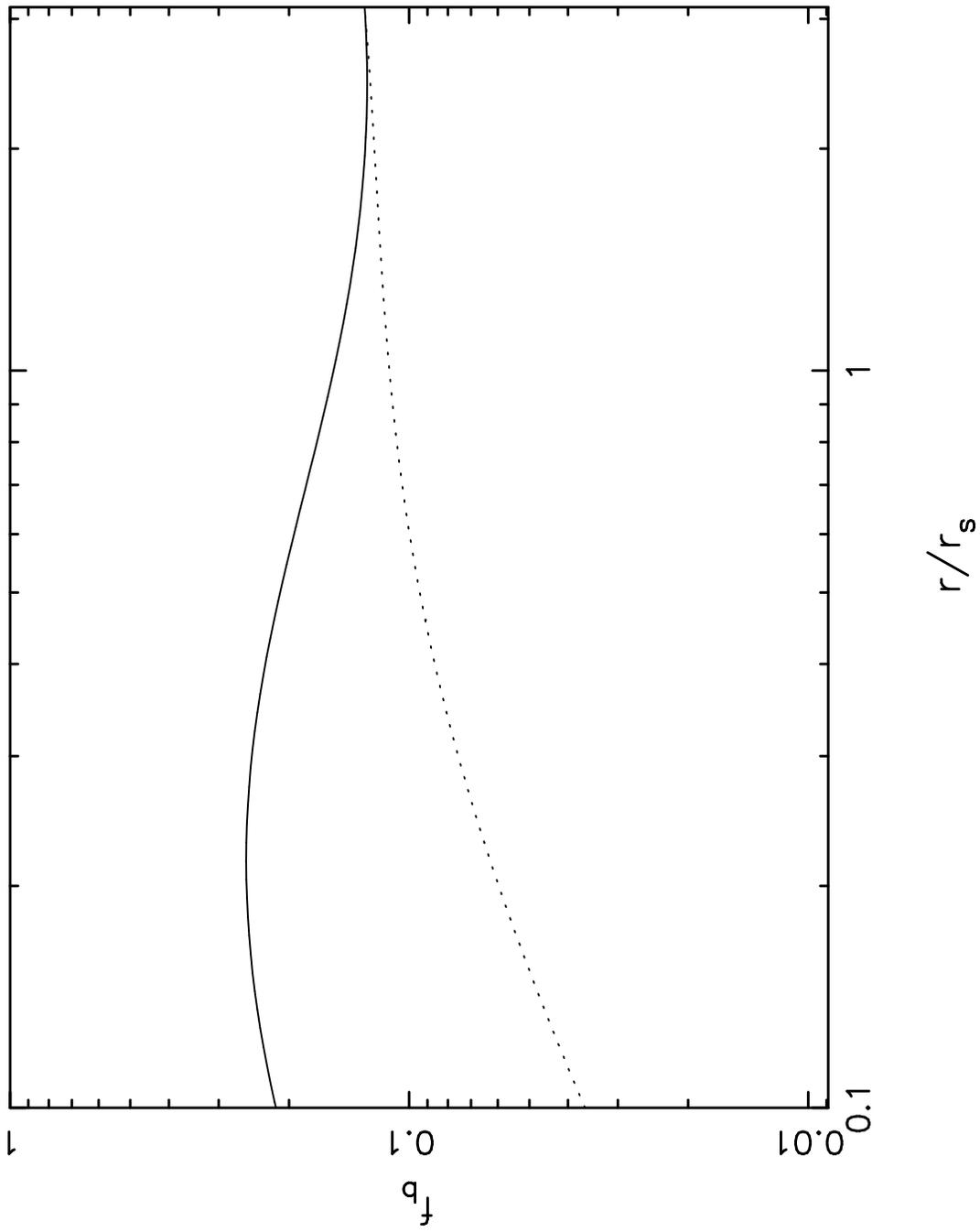
\end{document}